\documentclass[pra,aps,showpacs,reprint,superscriptaddress,twocolumn,amsmath,amssymb,nofootinbib,10pt]{revtex4-1}
  
\usepackage{graphicx,dcolumn,bm}
\usepackage{amsmath}
\usepackage{amsthm}
\usepackage{amsfonts}
\usepackage{mathrsfs}
\usepackage[usenames,dvipsnames]{xcolor}
\usepackage[colorlinks=true,citecolor=MidnightBlue,linkcolor=MidnightBlue,urlcolor=MidnightBlue]{hyperref}
\usepackage{braket}

\usepackage{bbm}

\newcommand{\average}[1]{\langle #1 \rangle}

\newcommand{\fr}{\mathbf{r}}

\newcommand{\G}[1]{G^{(#1)}}
\newcommand{\g}[1]{g^{(#1)}}

\renewcommand{\vec}[1]{{\bf #1}}

\newcommand {\startbild}{\begin{figure}}
\newcommand {\stopbild}{\end{figure}}
\newcommand {\staf}{\begin{equation}}
\newcommand {\stof}{\end{equation}}
\newcommand {\staffeld}{\begin{eqnarray}}
\newcommand {\stoffeld}{\end{eqnarray}}
\newcommand {\staa}{\begin{align}}
\newcommand {\stoa}{\end{align}}

\newcommand{\cref}[1]{Chapter~\ref{#1}}     
\newcommand{\eref}[1]{(\ref{#1})}           

\newcommand{\m}[1]{\left< #1 \right>}
\newcommand{\abs}[1]{\lvert #1 \rvert^2}

\newcommand{\hn}{\hat{n}}
\newcommand{\haa}{\hat{a}}

\begin{document}

\title{Simulating Dicke like superradiance with classical light sources}

\author{D.~Bhatti}
\affiliation{Institut f\"{u}r Optik, Information und Photonik, Universit\"{a}t Erlangen-N\"{u}rnberg, 91058 Erlangen, Germany}
\affiliation{Erlangen Graduate School in Advanced Optical Technologies (SAOT), Universit\"at Erlangen-N\"urnberg, 91052 Erlangen, Germany}

\author{S.~Oppel}
\affiliation{Institut f\"{u}r Optik, Information und Photonik, Universit\"{a}t Erlangen-N\"{u}rnberg, 91058 Erlangen, Germany}
\affiliation{Erlangen Graduate School in Advanced Optical Technologies (SAOT), Universit\"at Erlangen-N\"urnberg, 91052 Erlangen, Germany}

\author{R.~Wiegner}
\affiliation{Institut f\"{u}r Optik, Information und Photonik, Universit\"{a}t Erlangen-N\"{u}rnberg, 91058 Erlangen, Germany}

\author{G.~S.~Agarwal}
\affiliation{Erlangen Graduate School in Advanced Optical Technologies (SAOT), Universit\"at Erlangen-N\"urnberg, 91052 Erlangen, Germany}
\affiliation{Department of Physics, Oklahoma State University, Stillwater, Oklahoma 74078, USA}

\author{J.~von~Zanthier}
\affiliation{Institut f\"{u}r Optik, Information und Photonik, Universit\"{a}t Erlangen-N\"{u}rnberg, 91058 Erlangen, Germany}
\affiliation{Erlangen Graduate School in Advanced Optical Technologies (SAOT), Universit\"at Erlangen-N\"urnberg, 91052 Erlangen, Germany}

\date{\today}

\begin{abstract}
In this paper we investigate the close relationship between Dicke superradiance, originally predicted for an ensemble of two-level atoms in entangled states,
and the Hanbury Brown and Twiss effect, initially established in astronomy to determine the dimensions of classical light sources like stars. By studying the state evolution of the fields produced by classical sources -- defined by a positive Glauber-Sudarshan P function -- when recording intensity correlations of higher order 
in a generalized Hanbury Brown and Twiss setup we find that the angular distribution of the last detected photon, apart from an offset, is identical to the superradiant emission pattern generated by an ensemble of two-level atoms in entangled symmetric Dicke states. We show that the phenomenon derives from projective measurements induced by the measurement of photons in the far field of the sources and the permutative superposition of quantum paths identical to those leading to superradiance in the case of single photon emitters. We thus point out an important similarity between classical sources and quantum emitters upon detection of photons if the particular photon source remains unknown. We finally present a compact result for the characteristic functional which generates intensity correlations of arbitrary order for any kind of light sources.

\end{abstract}


\maketitle

\section{Introduction}

Superradiance is one of the enigmatic phenomena of quantum optics \cite{Dicke(1954),Agarwal(1970),Eberly(1971),Bonifacio(1971),Manassah(1973),AGARWAL(1974),Haroche(1982),Wiegner(2015),Longo(2016)}. In recent years, there has been considerable advance in the understanding of superradiance \cite{Scully(2006),Mazets(2007),Akkermans(2008),Scully(2009),Scully(2009)Super,Rohlsberger(2010),Wiegner(2011)PRA,Scully(2013),Rohlsberger(2013),Feng(2014),Scully(2015)PRL,Bhatti(2015),Longo(2016),Svidzinsky(2016)} and in bringing out new features, especially the statistical aspects of superradiance \cite{Jahnke(2015),Bhatti(2015),Jahnke(2016)}. For example, superradiance and subradiance have been understood to arise from the quantum interference of many paths which lead to a photon in the far zone \cite{Wiegner(2011)PRA}. The complexity of superradiance has been further revealed by studies focused on single photon \cite{Scully(2006),Scully(2007),Scully(2008),Scully(2009),Scully(2009)Super, Scully(2015),Rohlsberger(2010),Rohlsberger(2013)} or two photon excited \cite{Bhatti(2015)} superradiance, and also two atom superradiance in a cavity \cite{Reimann(2015),Casabone(2015)}. While the large body of literature has dealt with quantum emitters in line with the original work of Dicke \cite{Dicke(1954)}, one can ask if effects similar to Dicke's superradiance can be realized by much more commonly occurring sources like thermal sources \cite{Oppel(2014),Zhou(2015),Centini(2015)}. One of course has to keep in mind that not all the features would be simulated by the classical sources, i.e., sources defined by a positive
Glauber-Sudarshan P function \cite{Mandel(1995)}. Preliminary reports on the realization of specific features of superradiance with thermal sources have been published \cite{Oppel(2012),Oppel(2014)}. 

In a recent paper we demonstrated the isomorphism between Dicke superradiance, produced by atoms in highly entangled Dicke states, and higher order intensity correlation measurements of fields produced by initially uncorrelated single photon emitters (SPE) \cite{Oppel(2014),Wiegner(2015)}. From this analysis it became clear that the quantum mechanical measurement process, i.e., the state projection of the initially uncorrelated SPE onto particular Dicke states induced when spontaneously scattered photons are recorded in the far field of the sources, is the key for understanding the phenomenon. In this paper we show that this isomorphism holds also for classical light sources. In particular we demonstrate that (a) an equal mechanism of state projection derived for SPE in \cite{Oppel(2014),Wiegner(2015)} acts on fields produced by classical sources and that (b) this mechanism is at the origin of the superradiant emission pattern observed when measuring higher order intensity correlations in the far field of classical emitters.

The paper is organized as follows. In Sec.~\ref{sec:HBT_2TLS} we describe the essence of the scheme by first investigating the evolution of the field produced by \textit{two} classical light sources upon detection of photons in the far field. Here we demonstrate how via a Hanbury Brown and Twiss type of measurement involving two detectors Dicke superradiance is produced and explain the mechanism which leads to this result. In particular, we sort out that it is the state projection of the field occuring when the first photon is recorded which leads to the superradiant focused emission pattern observed when the second photon is detected. As a consequence, we demonstrate that the Hanbury Brown and Twiss effect, originally established in astronomy to determine the dimensions or distances of stars \cite{HBT(1956)Correlation,Brown(1967),HANBURY(1974)}, and Dicke superradiance, commonly observed with atoms in symmetric Dicke states \cite{Dicke(1954),Agarwal(1970),Eberly(1971),Bonifacio(1971),Manassah(1973),AGARWAL(1974),Haroche(1982)}, are two sides of the same coin.
We thereafter generalize the scheme to an arbitrary number of light sources and detectors. To that end we introduce in Sec.~\ref{sec:m-th order correlation function} the $m$th-order intensity correlation function and calculate it for an arbitrary number of sources $N$ with any kind of photon statistics. In Sec.~\ref{sec:Application} we apply the results of Sec.~\ref{sec:m-th order correlation function} to particular examples of classical and non-classical light sources. In Sec.~\ref{sec:Dicke} we then extend the idea of Sec.~\ref{sec:HBT_2TLS}, i.e., state projection of the field, to the superradiant $m$th-order correlation measurement leading to Dicke like states for classical sources. In Sec.~\ref{sec:functional} we present a compact way for calculating higher-order intensity correlations for any kind of light sources based on the  characteristic functional. In Sec.~\ref{sec:conclusion} we present our concluding remarks.

\section{Dicke like superradiance and Hanbury Brown and Twiss effect for two classical light sources}
\label{sec:HBT_2TLS}

\begin{figure}[b]
\centering
\includegraphics[width=0.45\textwidth]{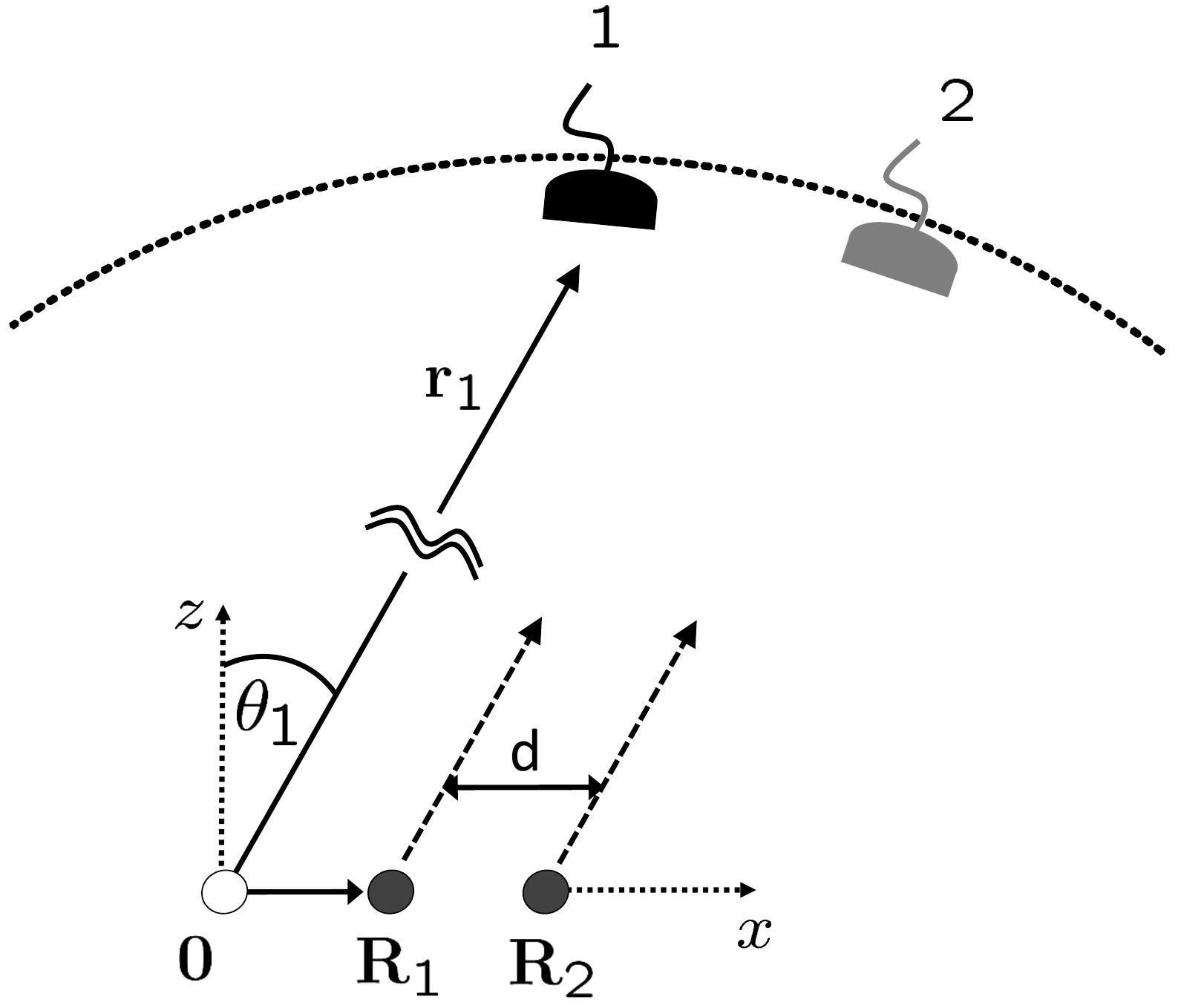}
\caption{Considered setup of Sec.~\ref{sec:HBT_2TLS}: Two identical classical light sources, separated by a distance $d \gg \lambda$, are placed at positions $\vec{R}_{l}$, $l = 1, 2$; the light scattered by the sources is measured by two detectors, located at positions $\vec{r}_j$, $j = 1, 2$, in the far field of the sources.}
\label{fig:detection_scheme}
\end{figure}

In this section we consider two classical light sources located at ${\vec R}_{1}$ and ${\vec R}_{2}$, separated by a distance $d$ much larger than the wavelength $\lambda$ of the emitted photons so that any direct interaction of the sources can be neglected (see Fig.~\ref{fig:detection_scheme}). 
The quantity of interest is the spatial intensity correlation $\left< I(\vec{r}_{1})I(\vec{r}_{2}) \right>$ obtained when correlating the intensities $I(\vec{r}_{1})$ and $I(\vec{r}_{2})$ in the far field of the sources at the two positions $\vec{r}_{1}$ and $\vec{r}_{2}$, respectively. 
The far field condition is crucial in the setup as it ensures the indistinguishability of the recorded photons, i.e., upon detection of a photon it is principally impossible to identify the individual photon source. To simplify the calculations we assume the detectors to be placed in one plane with the sources and arranged in a circle around the emitters (see Fig.~\ref{fig:detection_scheme}). Note that in the case of thermal light sources (TLS) this arrangement corresponds to the original Hanbury Brown and Twiss setup
to determine the diameter or distances of stars \cite{HBT(1956)Correlation,Brown(1967),HANBURY(1974)}; for two SPE this setup has been investigated in \cite{Oppel(2014),Wiegner(2015)}.

The first and second order intensity correlation functions for an arbitrary light field $\rho$ are defined as \cite{Glauber(1963)Quantum,Glauber(1963)Coherent}
\begin{eqnarray}
\label{eq:G2}
G^{(1)}_{\rho}(\vec{r}) & = & \Big< E^{(-)}(\vec{r})E^{(+)}(\vec{r})) \Big>_{\rho} \\
	G^{(2)}_{\rho}(\vec{r}_{1},\vec{r}_{2}) & = & \Big< E^{(-)}(\vec{r}_{2})E^{(-)}(\vec{r}_{1})E^{(+)}(\vec{r}_{1})E^{(+)}(\vec{r}_{2}) \Big>_{\rho} \nonumber 
\end{eqnarray}
where the positive and negative frequency parts of the electric field operator, $E^{(+)}(\vec{r})$ and $E^{(-)}(\vec{r})$, respectively, due to the far field condition take the form \cite{Wiegner(2015)}
\begin{equation}
\left[ E^{(-)}(\vec{r}) \right]^{\dagger}	= E^{(+)}(\vec{r}) =\sum_{l=1}^{2} e^{i\frac{\omega}{c}\vec{n}\cdot \vec{R}_{l}} \hat{a}_{l} \, .
\label{eq:Eplus}
\end{equation}
In Eq.~(\ref{eq:Eplus}), $\hat{a}_{l}$ defines the annihilation operator of a photon from source $l$, $\omega/c = 2 \pi/\lambda = k$, and $\vec{n}=\vec{r}/|\vec{r}|$ is the direction of propagation of a photon recorded at $\vec{r}$.
Note that for simplicity we define the field and hence all correlation functions dimensionless; the actual values can be obtained by multiplying $G^{(m)}_{\rho}$ with $2m$ times the electric field amplitude $\mathcal{E}_{0}$ of a single source.

The field  $\rho_{2}$ produced by $N=2$ light sources can be written in the photon number basis as
\begin{equation}
	\rho_{2} =  \sum_{n_{1},n_{2}=0}^{\infty} P(n_{1})P(n_{2}) \ket{n_{1},n_{2}}\bra{n_{1},n_{2}} \ ,
\label{eq:rho_2TLS}
\end{equation}
where $P(n_{l})$ describes the photon statistics of source $l$, $l=1, 2$, and $\ket{n_{1},n_{2}}\bra{n_{1},n_{2}}$ is the tensor product of the number-state vectors $\ket{n_{1}}\bra{n_{1}}\otimes \ket{n_{2}}\bra{n_{2}}$. Note that for TLS the photon statistics $P(n_{l})$ is given by a Bose-Einstein distribution \cite{SCULLY(1997)}
\begin{equation}
  \label{distributionthermal}
  P_{TLS}(n_l)=\frac{1}{1+\bar{n}_l} \left( \frac{\bar{n}_l}{1+\bar{n}_l}\right)^{n_l} \, ,
\end{equation}
where $\bar{n}_{l}$ denotes the mean photon number of source $l$ given by
\begin{equation}
  \label{measuring}
  \bar{n}_{l} = \braket{\hat{n}_{l} }_{\rho} = \braket{\hat{a}_{l}^{\dagger}\hat{a}_{l}}_{\rho} \ .
\end{equation}

Using Eqs.~(\ref{eq:Eplus}) - (\ref{measuring}) in Eq.~(\ref{eq:G2}) and assuming $\bar{n}_{l}=\bar{n}$ to be identical for the two sources, the normalized second order intensity correlation function for two TLS calculates to
\begin{equation}
\begin{aligned}
	g^{(2)}_{\rho_{2 \, TLS}}(\vec{r}_{1},\vec{r}_{2})&= \frac{G_{\rho_{2 \, TLS}}^{(2)}(\vec{r}_{1},\vec{r}_{2})}{G_{\rho_{2 \, TLS}}^{(1)}(\vec{r}_{1})G_{\rho_{2 \, TLS}}^{(1)}(\vec{r}_{2})} \\
	 &= \frac{3}{2} [ 1 + \frac{1}{3} \cos\left(k \, (\vec{n}_{1}-\vec{n}_{2})\cdot (\vec{R}_{1}-\vec{R}_{2})\right) ] \ , 
\label{eq:g2}
\end{aligned}
\end{equation}
where
\begin{equation}		
		G^{(1)}_{\rho_{2 \, TLS}}(\vec{r}) = \Big< E^{(-)}(\vec{r}) E^{(+)}(\vec{r}) \Big>_{\rho_{2 \, TLS}} = 2 \bar{n} \ .
\label{eq:G1}
\end{equation}
The result of Eq.~(\ref{eq:g2}) is well-known from the original Hanbury Brown and Twiss measurement for a double star system \cite{Brown(1967),HANBURY(1974)}. It moreover shows similarities to the normalized second order intensity correlation function obtained for a field produced by two uncorrelated SPE, e.g., two two-level atoms in the fully excited state $\ket{e,e}$ \cite{Wiegner(2015)}
\begin{equation}
\begin{aligned}
	g^{(2)}_{\rho_{2 \, SPE}}(\vec{r}_{1},\vec{r}_{2})= \frac{1}{2} [ 1 + \cos\left(k \, (\vec{n}_{1}-\vec{n}_{2})\cdot (\vec{R}_{1}-\vec{R}_{2})\right) ] \ .
\label{eq:g2atoms}
\end{aligned}
\end{equation}
Both distributions, Eqs.~(\ref{eq:g2}) and (\ref{eq:g2atoms}), display identical modulations with respect to the detector positions at $\vec{r}_{1}$ and $\vec{r}_{2}$, indicating that the distributions derive from the same interference phenomenon. Note, however, that the second order intensity correlation function for two TLS, Eq.~(\ref{eq:g2}), is bound from below, i.e., $g^{(2)}_{2 \, TLS}(\vec{r}_{1},\vec{r}_{2}) \geq 1$, whereas for two SPE $g^{(2)}_{2 \, SPE}(\vec{r}_{1},\vec{r}_{2})$ vanishes at $k (\vec{n}_{1}-\vec{n}_{2})\cdot(\vec{R}_{1}-\vec{R}_{2}) = \pi$ (see Eq.~(\ref{eq:g2atoms})). This means that for two TLS, unlike for two SPE, the joint probability of detecting one photon at $\vec{r}_{2}$ and another photon at $\vec{r}_{1}$ never drops to zero.

For two SPE the interference pattern of Eq.~(\ref{eq:g2atoms}) has been identified as Dicke super- and subradiance \cite{Oppel(2014),Wiegner(2015)}. This is due to the fact that the measurement of the first photon, depending on the point of detection ${\vec r}_{1}$, projects the two initially uncorrelated atoms in the state $\ket{e,e}$ onto the symmetric Dicke state $\ket{+}$ (antisymmetric Dicke state $\ket{-}$), given by $\ket{\pm}= \frac{1}{\sqrt{2}} \left( \ket{e,g} \pm \ket{g,e} \right)$ \cite{Wiegner(2015)}, where $\ket{g}$ ($\ket{e}$) denotes the ground (excited) state of a single atom. 

We next show that the second order correlation function for the radiation produced by two independent TLS can be interpreted as the radiation pattern produced by two correlated sources. For this porpuse we write the second order correlation function in the form \cite{Wiegner(2015)}
\begin{equation}
\begin{aligned}
	G_{\rho_{2}}^{(2)}(\vec{r}_{1},\vec{r}_{2}) &= \text{Tr}[\rho_{2} \, E^{(-)}(\vec{r}_{1})E^{(-)}(\vec{r}_{2})E^{(+)}(\vec{r}_{2})E^{(+)}(\vec{r}_{1})] \\
	& = G_{\tilde{\rho}_{2}}^{(1)}(\vec{r}_{2}) G_{\rho_{2}}^{(1)}(\vec{r}_{1}) \ ,
\label{eq:G2_rho_G1}
\end{aligned}
\end{equation}
with
\begin{equation}
G_{\tilde{\rho}_{2}}^{(1)}(\vec{r}_{2}) = \text{Tr}\left[\tilde{\rho_{2}}  E^{(-)}(\vec{r}_{2})E^{(+)}(\vec{r}_{2})\right] \ .
\label{eq:G_rho_2_tilde}
\end{equation}
This shows that also for arbitrary classical fields $\rho_{2}$ we find an isomorphism between $G_{\rho_{2}}^{(2)}(\vec{r}_{2},\vec{r}_{1})$ and $G_{\tilde{\rho}_{2}}^{(1)}(\vec{r}_{2})$
, where
\begin{equation}
\tilde{\rho}_{2} = \frac{E^{(+)}(\vec{r}_{1}) \rho_{2} E^{(-)}(\vec{r}_{1})}{\text{Tr}[ \rho_{2} E^{(-)}(\vec{r}_{1})E^{(+)}(\vec{r}_{1}) ]} \ \ , \ \text{Tr}[\tilde{\rho}_{2}] = 1 \, . 
\label{eq:rho_tilde}
\end{equation}
Note that $E^{(+)}(\vec{r}_{1})$ is the annihilation operator and hence $\tilde{\rho}_{2}$ is obtained by subtracting a photon from the initial field $\rho_{2}$.

Since we assume in Fig.~\ref{fig:detection_scheme} the two classical sources to be initially uncorrelated the first order intensity correlation function does not display 
any modulation (cf., e.g., Eq.~(\ref{eq:G1})). However, when performing an intensity measurement after subtracting a photon from the state $\rho_{2}$, i.e., measuring $G_{\tilde{\rho}_{2}}^{(1)}(\vec{r}_{2})$ of the projected state $\tilde{\rho}_{2}$, 
interference fringes are observed (cf. Eqs.~(\ref{eq:g2}) and (\ref{eq:G2_rho_G1}) - (\ref{eq:rho_tilde})). These interferences result from the correlations between the two sources induced by the detection of the first photon at $\vec{r}_{1}$. 

Indeed, the state of the field after the detection of the first photon reads (cf. Eq.~(\ref{eq:rho_tilde}))
\begin{equation}
\begin{aligned}
	\tilde{\rho}_{2} & =  \frac{E^{(+)}(\vec{r}_{1}) \rho_{2} E^{(-)}(\vec{r}_{1})}{\text{Tr}[\rho_{2} E^{(-)}(\vec{r}_{1})E^{(+)}(\vec{r}_{1})]} \\
		& = \frac{1}{2\bar{n}} \big[ a_{1} \rho_{2} a_{1}^{\dagger} + a_{2} \rho_{2} a_{2}^{\dagger} 
		\\  & \hphantom{=} + e^{i k \vec{n}_{1}(\vec{R}_{1}-\vec{R}_{2})} a_{1} \rho_{2} a_{2}^{\dagger} + e^{-i k \vec{n}_{1}(\vec{R}_{1}-\vec{R}_{2})} a_{2} \rho_{2} a_{1}^{\dagger} \big] \ ,
\label{eq:rho_2TLS_proj}
\end{aligned}
\end{equation}
which is not of a diagonal form. The non-diagonal terms are highlighted by the acquired mode-mode correlation,
i.e., by
\begin{equation}
\text{Tr}[\tilde{\rho_{2}}a_{1}^{\dagger}a_{2}] = \frac{\bar{n}}{2} e^{ik\vec{n}_{1}(\vec{R}_{1}-\vec{R}_{2})} \ .
\label{eq:correlation}
\end{equation}
Clearly $\tilde{\rho}_{2}$ can be interpreted as the density matrix of the sources after the detection of a photon at $\vec{r}_{1}$ and the non-diagonal terms produce the correlations between the two sources. This mode-mode correlation leads to the fringe pattern in Eq.~(\ref{eq:G_rho_2_tilde}). Note that the degree of correlation (Eq.~(\ref{eq:correlation})) depends on the position where the first photon is recorded.

Equally to the case of two SPE in the Dicke state $\ket{\pm}$
the correlated \textit{classical Dicke state} Eq.~(\ref{eq:rho_2TLS_proj}) emits super- or subradiant light. This is caused by the same permutative superposition of quantum paths upon detection of the last photon, i.e., the same kind of constructive and destructive interferences among multiple photon pathways, as those leading to superradiance in the case of SPE \cite{Wiegner(2011)PRA,Oppel(2014)}. This is discussed in the next section.

\section{\texorpdfstring{$m$}{}th-order intensity correlation function for arbitrary light sources}
\label{sec:m-th order correlation function}

In this section we investigate intensity correlations of order $m \geq 1$ for arbitrary light fields $\rho_N$ produced by $N$ identical sources with arbitrary photon statistics. 
Note that higher-order intensity correlations of light fields produced by classical sources have been formerly studied, both theoretically and experimentally \cite{Agafonov(2008),Liu(2009),Iskhakov(2011),Chen(2011)}, leading to new insights into, e.g., ghost imaging \cite{Chan(2009),Agafonov(2009)arxiv,Zhou(2010),Chen(2010),Erkmen(2010)}, quantum imaging \cite{Oppel(2012),Oppel(2014)}, or quantum information processing  \cite{Tamma(2014),Tamma(2015)}. 

\begin{figure}[b]
\centering
\includegraphics[width=0.45\textwidth]{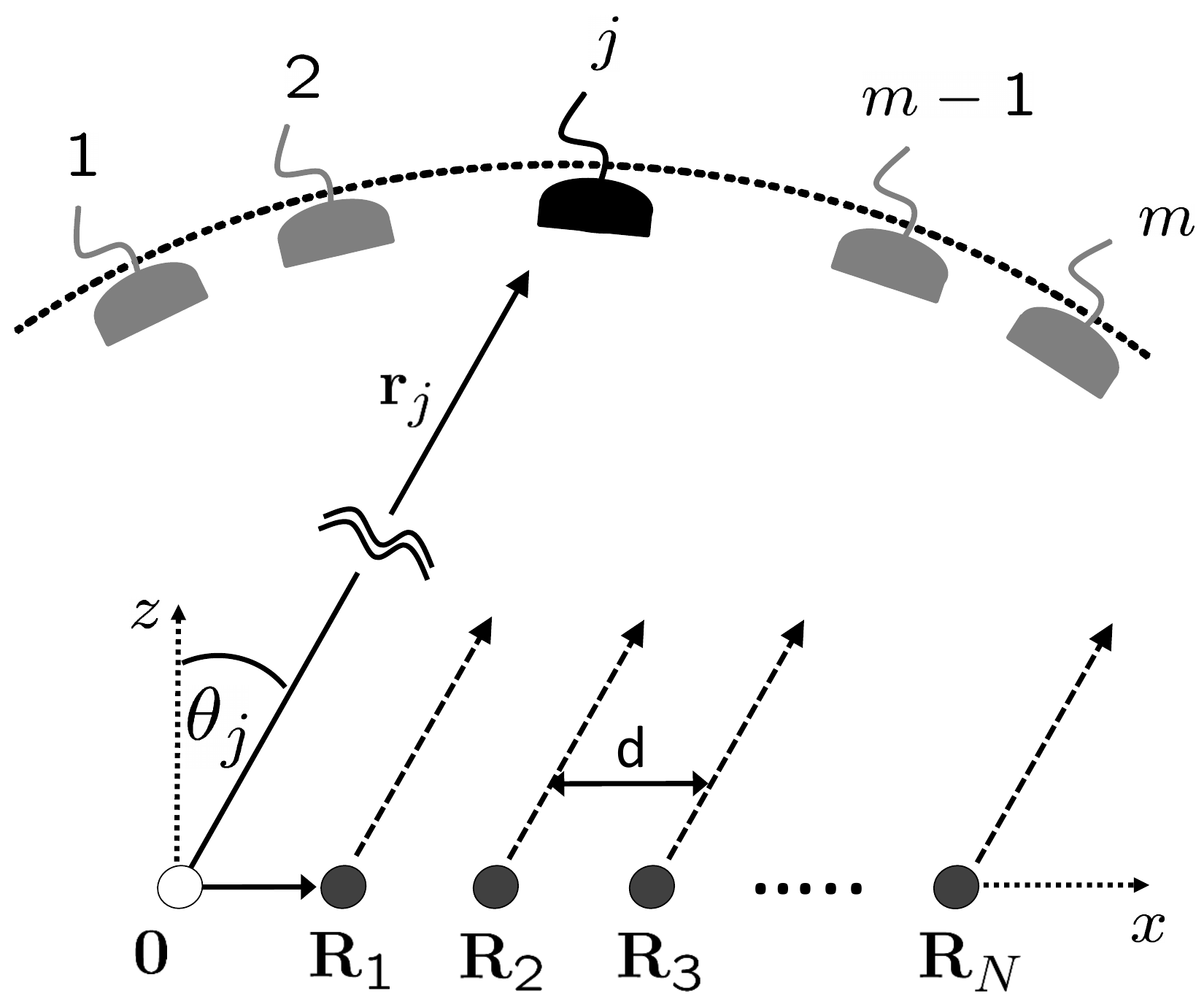}
\caption{Considered setup of Sec.~\ref{sec:m-th order correlation function}: $N$ identical arbitrary light sources, separated by an equal distance $d \gg \lambda$, are placed along a chain at positions ${\vec R}_{l}$, $l = 1, \ldots , N$; the light scattered by the sources is measured by $m$ detectors, located at positions ${\vec r}_j$, $j = 1, \ldots , m$, in the far field of the sources.}
\label{fig:scheme_N}
\end{figure}

To simplify the following calculations we assume the sources to be aligned along a chain at positions $\vec{R}_{l}$, $l=1,\hdots,N$, and separated by equal distances $d \gg \lambda$ so that any direct interaction of the sources can be neglected (see Fig.~\ref{fig:scheme_N}). Again, the $m$ detectors measuring the intensities $I(\vec{r}_{j})$ at positions $\vec{r}_{j}$, $j=1,\hdots, m$, are assumed to be placed in one plane in a circle around the sources. Moreover, the detectors have to be located in the far field of the sources to ensure the indistinguishability of the recorded photons. For $N$ sources the $m$th-order intensity correlation function reads \cite{Glauber(1963)Quantum,Glauber(1963)Coherent}
\begin{equation}
\label{Eq1n}
G^{(m)}_{N} (\vec{r}_1, \hdots , \vec{r}_m)  = \average{:\prod_{j=1}^m E^{(-)}(\fr_j) E^{(+)}(\fr_j):}_{\rho_N} \, ,
\end{equation}
where $\average{: \ldots :}_{\rho_N}$ denotes the normally ordered quantum mechanical expectation value for a field in the state $\rho_N$ and the operators $E^{(+)}(\vec{r}_j) = \left[ E^{(-)}(\vec{r}_j) \right]^{\dagger}$, $j=1,\hdots, m$, are defined as in Eq.~(\ref{eq:Eplus}). 

Writing as in Eq.~(\ref{eq:rho_2TLS}) the density matrix of the field as ${\rho_N} = \sum_i P_i \ket{\psi_i}\bra{\psi_i}$, where $P_i$ is the probability to find the field in the state $\ket{\psi_i}$, we obtain in the number state representation
\begin{equation}
\label{eq:rho_N-sources}
\begin{aligned}
{\rho_N} &= \sum_{i = 1}^N P_i \ket{\psi_i}\bra{\psi_i} \\
& = \sum_{n_1,\ldots,n_N=0}^{\infty} P(n_1)P(n_2) \ldots P(n_N) \\
& \hspace{10mm}  \times \ket{n_1,n_2,\ldots,n_N} \bra{n_1,n_2,\ldots,n_N} \, .
\end{aligned}
\end{equation}
Plugging Eq.~(\ref{eq:rho_N-sources}) into Eq.~(\ref{Eq1n}) and using Eq.~(\ref{eq:Eplus}) we obtain for $m-1$ fixed detectors at the same position $\vec{r}_1$ as a function of the $m$th detector at $\vec{r}_2$ (see \cite{Oppel(2014),Wiegner(2015)})
\begin{equation}
\begin{aligned}
  & G^{(m)}_{N}  (\vec{r}_1, ...,\vec{r}_1,\vec{r}_2)  \\
	& = \sum_{n_1,\ldots,n_N=0}^{\infty} P(n_1)P(n_2) \ldots P(n_N) \sum_{\{n_l\}} \Big| \bra{\{n_l\}} \\
	& \, \times \left(\sum_{l=1}^{N}e^{i\,\varphi_{l1}}\haa_l\right)^{m-1}\!\! \left(\sum_{l=1}^{N}e^{i\,\varphi_{l2}}\haa_l\right) \ket{n_1,n_2,\ldots,n_N}\Big|^{2} \, ,
 \label{gn_2_foc_10}
\end{aligned}
\end{equation}
where $\bra{\{n_{l}\}}$ denotes all orthonormal multi-mode eigenstates of the system in the number state representation and 
\begin{equation}
\label{eq:varphi_lj}
\varphi_{lj}  = k \, \frac{\vec{R}_l \cdot \vec{r}_j}{r_j} = l\,kd\,\sin\theta_j \ ,
\end{equation}
equals the optical phase accumulated by a photon emitted at $\vec{R}_l$ and detected at ${\vec r}_j$ relative to a photon emitted at the origin (cf.~Fig.~\ref{fig:scheme_N}).

Exploiting the orthogonality of the number states and using the multinomial formula $(x_1+x_2+\ldots+x_N)^{m} = \sum_{m_1+m_2+\ldots+m_N=m}\binom{m}{m_1,m_2,\ldots,m_N}x_1^{m_1}x_2^{m_2} \ldots x_N^{m_N}$ we can write Eq.~(\ref{gn_2_foc_10}) also in the form
\begin{widetext}
\begin{align}
  \label{gn_2_foc_11}
  & G^{(m)}_{N} (\vec{r}_1, ...,\vec{r}_1,\vec{r}_2)  = \sum_{n_1,\ldots,n_N=0}^{\infty} P(n_1)P(n_2) \ldots P(n_N) \nonumber\\
  & \hspace{50pt}\times \sum_{m_l} \abs{\bra{n_1-m_1,n_2-m_2,\ldots,n_N-m_N} \haa_1^{m_1}\haa_2^{m_2}\ldots\haa_N^{m_N} \ket{n_1,n_2,\ldots,n_N}} \nonumber\\
  & \hspace{50pt}\times\left|
    \binom{m-1}{m_1-1,m_2,\ldots,m_N}e^{i[\varphi_{12}+(m_1-1)\varphi_{11}+m_2\varphi_{21}+\ldots+m_N\varphi_{N1}]} \right. \nonumber\\ 
  &\hspace{60pt}+\binom{m-1}{m_1,m_2-1,\ldots,m_N}e^{i[\varphi_{22}+m_1\varphi_{11}+(m_2-1)\varphi_{21}+\ldots+m_N\varphi_{N1}]}   \nonumber\\ 
  &\hspace{70pt}+\ldots \nonumber\\ 
  &\hspace{80pt}+\left.\binom{m-1}{m_1,m_2,\ldots,m_N-1}e^{i[\varphi_{N2}+m_1\varphi_{11}+m_2\varphi_{21}+\ldots+(m_N-1)\varphi_{N1}]} \right|^2 \, ,
\end{align}
\end{widetext}
where $\binom{m}{m_1,m_2,\ldots,m_N}=\frac{m!}{m_1!m_2!\ldots m_N!}$ denotes the multinomial coefficient and $m_l$ is the number of photons emitted by the $l$th source. Note that in Eq.~\eref{gn_2_foc_11} the expression $\sum_{m_l}\equiv\sum_{m_1+m_2+\ldots+m_N=m}$ runs over all combinations of integers $m_1$ through $m_N$ in such a way that $m_1+m_2+\ldots+m_N=m$. Here, the number of different combinations, i.e., of different realizations of the sum $m_1+m_2+\ldots+m_N=m$, defines the number of final states which appear in $G^{(m)}_{N} (\vec{r}_1, ...,\vec{r}_1,\vec{r}_2)$. Hence, the $m$th-order correlation function $G^{(m)}_{N} (\vec{r}_1, ...,\vec{r}_1,\vec{r}_2)$ is the incoherent sum of $\binom{N+m-1}{m}$ different terms resulting from $\binom{N+m-1}{m}$ different final states, where each final state gives rise to an individual sub-interference pattern generated by the coherent superposition of $\binom{m}{m_1,m_2,\ldots,m_N}$ indistinguishable yet different $m$-photon quantum paths leading to the same final state (cf. Eq.~\eref{gn_2_foc_11}). 
Summing over all final states we obtain the total number of indistinguishable yet different $m$-photon quantum paths contributing to $G^{(m)}_{N} (\vec{r}_1, ...,\vec{r}_1,\vec{r}_2)$, what yields $\sum_{m_l}\binom{m}{m_1,m_2,\ldots,m_N} = N^m$. 

We can simplify Eq.~\eref{gn_2_foc_11} further if we factor out the multinomial coefficient $\binom{m}{m_1,m_2,\ldots,m_N}$ and rearrange the complex phase terms. We thus obtain
\begin{align}
  \label{gn_2_foc_12}
  & G^{(m)}_{N} (\vec{r}_1, ...,\vec{r}_1,\vec{r}_2)  \nonumber\\ 
  & = \frac{1}{m^2}\sum_{m_l}\left[\prod_{l=1}^{N} \m{:\hn_l^{m_l}:}_{{\rho_{N}}}\binom{m}{m_1,m_{2},\ldots,m_N}^2  \right. \nonumber\\
  & \hspace{40mm}\times \left. \Big|\sum_{l'=1}^{N} m_{l'} e^{-i\, l' \varphi_{\Delta}} \Big|^2\right] \, ,
\end{align}
where we introduced the relative phase
\staf
\label{varphidelta}
\varphi_{\Delta} \equiv \frac{ \varphi_{l' 1} - \varphi_{l' 2} }{l'}= \varphi_{11} - \varphi_{12} \; .
\stof
From Eqs.~\eref{gn_2_foc_12} and \eref{varphidelta} we can see that for each final state the effective interference term arises from the emission of the last photon recorded by the detector at $\vec{r}_2$, where the number $m_{l'}$, $l'=1,\ldots,N$, in Eq.~\eref{gn_2_foc_12} denotes the number of events for which the last photon is emitted by source $l'$.
Note that in the case of $m_{j}=m$ and $m_{i\neq j}=0$ there is only one possible $m$-photon quantum path so that in this particular case no interference term appears; this configuration is responsible for an offset in Eq.~\eref{gn_2_foc_12} and emerges only if the sources are not SPE (compare Eqs.~(\ref{eq:g2}) and (\ref{eq:g2atoms})).

Let us investigate the various interference terms appearing in Eq.~(\ref{gn_2_foc_12}) in more detail. To that aim we write the sum over all final states $\sum_{m_{l}}$, running over all combinations of integers $m_{1}$ through $m_{N}$ such that $m_{1}+\ldots+m_{N}=m$, in terms of the partitions of the number $m$, i.e., all combinations of integers $x_{1} \leq x_{2} \leq \ldots \leq x_{N}$, so that $x_{1}+\ldots+x_{N}=m$. The sum over all final states then takes the form
\begin{equation}
	\sum_{m_{l}} = \sum_{\substack{ x_{l} \\  x_{1} \leq x_{2} \leq \ldots \leq x_{N} }} \sum_{\{m_{l}\}\in S_{\{x_{l}\}}} \ ,
\label{eq:sum_partitions}
\end{equation}
where $S_{\{x_{l}\}}$ denotes the symmetric group of elements $\{x_{l}\}=\{x_{1},\ldots,x_{N}\}$. Note that the first sum on the right hand side of Eq.~(\ref{eq:sum_partitions}) runs over all partitions of the integer $m$, whereas the second sum lists all permutations of a given set $\{x_{l}\}$ among all sources corresponding to all final states of a given partition.  

\begin{figure}[t]
\centering
\includegraphics[width=0.45\textwidth]{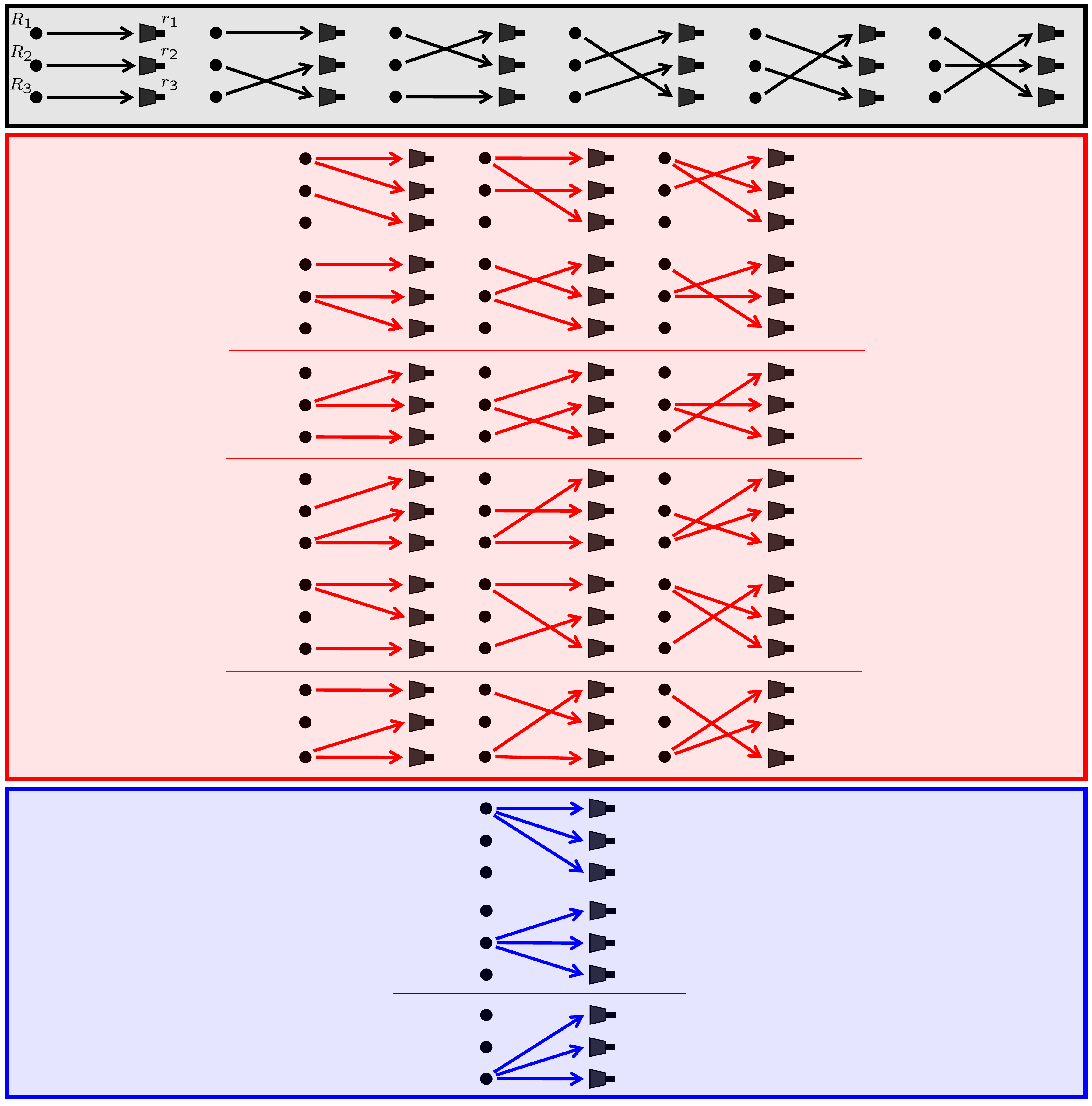}
\caption{(Color online) All $27$ three-photon quantum paths contributing to the third-order correlation function $G^{(3)}_{3} (\vec{r}_1,\vec{r}_2,\vec{r}_3)$ in the case of $N=3$ identical classical light sources at different positions $\vec R_l$ ($l=1,2,3$). The three boxes (in the three colors gray, red, and blue) represent the three different partitions of the number $3$, i.e., the  different possibilities of emitting three photons by three sources: upper box (gray) lists all three-photon quantum paths where each source emits exactly one photon (partition $\{1,1,1\}$); middle box (red) contains all three-photon quantum paths where one source emits one photon and another source two photons (partition $\{0,1,2\}$); lower box (blue) lists all three-photon quantum paths where all three photons stem from the same source (partition $\{0,0,3\}$). Each line in the figure lists the three-photon emission processes resulting in the same final state. Note that for the superradiant condition $\vec{r}_1 = \vec{r}_2$ all three-photon quantum paths within a given partition are produced fully symmetrically (cf. Eq.~\eref{eq:Partitions_Superradiance}).}
\label{fig:quantum_paths}
\end{figure}

Since the sources are assumed to be identical the product of the statistical moments $\prod_{l=1}^{N} \m{:\hn_l^{m_l}:}_{{\rho_{N}}}$ is equal for all permutations within a given partition. The same is true for the multinomial coefficient and thus, by use of Eq.~(\ref{eq:sum_partitions}), we can rewrite Eq.~(\ref{gn_2_foc_12}) also in the form
\begin{equation}
\begin{aligned}
  & G^{(m)}_{N} (\vec{r}_1, ...,\vec{r}_1,\vec{r}_2)  \\ 
	& = \frac{1}{m^2}   \sum_{\substack{ x_{l} \\  x_{1} \leq x_{2} \leq \ldots \leq x_{N} }}  \prod_{l=1}^{N} \m{:\hn_l^{x_l}:}_{{\rho_{N}}}\binom{m}{x_1,x_{2},\ldots,x_N}^2   \\
  & \hspace{30mm}\times \sum_{\{m_{l}\}\in S_{\{x_{l}\}}} \Big|\sum_{l'=1}^{N} m_{l'} e^{-i\, l' \varphi_{\Delta}} \Big|^2 \, .
\label{eq:GNpartitions}
\end{aligned}
\end{equation}
As can be seen from Eq.~(\ref{eq:GNpartitions}), the interference term within a given partition derives from the sum over all final states within that partition. The latter calculates to
\begin{equation}
\begin{aligned}
&\sum_{\{m_{l}\}\in S_{\{x_{l}\}}} \Big|\sum_{l'=1}^{N} m_{l'} e^{-i \, l' \varphi_{\Delta}} \Big|^2 \\
&= \sum_{\{m_{l}\}\in S_{\{x_{l}\}}} \left[ \sum_{l'=1}^{N}m_{l'}^{2} +  \sum_{\substack{ l',l''=1 \\  l' \neq l '' }  }^{N} m_{l'}m_{l''} e^{-i\, l' \varphi_{ \Delta}}e^{+i\, l'' \varphi_{ \Delta}} \right] \\
&=  c_{1} + c_{2} \left[  \frac{\sin^{2} \left(  N \frac{\varphi_{\Delta}}{2} \right)}{\sin^{2} \left(  \frac{\varphi_{\Delta}}{2} \right)} - N \right] \, ,
\label{eq:Partitions_Superradiance}
\end{aligned}
\end{equation}
where the last step in Eq.~(\ref{eq:Partitions_Superradiance}) is the result of the symmetric occurrence of final states within a given partition, 
i.e., the symmetric permutation of a given set $\{x_l\}$ among all sources (see Fig.~\ref{fig:quantum_paths}). 
This permutation is the quintessence of spatial superradiance as it leads for each weighting factor $x_{k'}x_{k''}$ of a given set $\{x_l\}$, $k',k''= 1, \ldots, N$ ($k' \neq k''$), to the same superposition of all possible relative phase terms $e^{-i\, l' \varphi_{ \Delta}}e^{+i\, l'' \varphi_{ \Delta}}$, $l',l''= 1, \ldots, N$ ($l' \neq l''$), giving rise to the peaked emission pattern of Eq.~\eref{eq:Partitions_Superradiance}.

As can be seen from Eq.~(\ref{eq:Partitions_Superradiance}), the phenomenon of superradiance is linked to the principal impossibility to identify the individual photon source upon detection of a photon. This is ensured by the far field configuration assumed in our setup (see Figs.~\ref{fig:detection_scheme} and \ref{fig:scheme_N}). Moreover, Eq.~(\ref{eq:Partitions_Superradiance}) reveals that superradiance arises for each partition separately. Thus, summing up all partitions to calculate $G^{(m)}_{N} (\vec{r}_1, ...,\vec{r}_1,\vec{r}_2)$, each weighted by its corresponding statistics (cf. Eq.~(\ref{eq:GNpartitions})), we obtain the same focused superradiant spatial emission pattern as displayed by Eq.~(\ref{eq:Partitions_Superradiance}) alone. Note that such a peaked angular distribution as a function of $\vec{r}_2$ produced by incoherent sources was known so far only to occur for SPE \cite{Dicke(1954),Eberly(1971),Manassah(1973),Haroche(1982),AGARWAL(1974)}. However, as Eq.~(\ref{eq:GNpartitions}) shows, the case of SPE represents merely a particular realisation of $G^{(m)}_{N} (\vec{r}_1, ...,\vec{r}_1,\vec{r}_2)$, resulting from the single partition $\{1,\hdots,1\}$ \cite{Wiegner(2015)}. The appearance of superradiance for other kinds of light sources will be discussed in the next section.

\section{Application to particular examples}
\label{sec:Application}

In this section we apply the general outcome for the $m$th-order correlation function $G^{(m)}_{N} (\vec{r}_1, ...,\vec{r}_1,\vec{r}_2)$, Eqs.~\eref{gn_2_foc_12} and \eref{eq:GNpartitions}, to particular examples. Note that in Eqs.~\eref{gn_2_foc_12} and \eref{eq:GNpartitions} we have not yet made any assumptions about the photon statistics of the light sources considered and the given expressions are thus generally valid for any kind of sources.

Considering for example $N$ identical initially uncorrelated TLS with equal mean photon numbers $\bar{n}_{l} = \bar{n}$, we find 
\begin{align}
  \label{gn_2_foc_15}
  & \G{m}_{N\,TLS}(\vec{r}_1, ...,\vec{r}_1,\vec{r}_2)  \nonumber\\
  &  = \bar{n}^m N^{m}(m-1)!\left(1 + \frac{m-1}{N^2}\frac{\sin^2{\left(N\frac{\varphi_{11}-\varphi_{12}}{2}\right)}}{\sin^2{\left(\frac{\varphi_{11}-\varphi_{12}}{2}\right)}}\right) \, ,
\end{align}
\noindent where in order to derive Eq.~\eref{gn_2_foc_15} we took advantage of the multinomial identity
\begin{align}
   & \quad \sum_{m_l}\binom{m}{m_1,m_2,\ldots,m_N} m_k m_{k'} = \nonumber\\
   & \begin{cases}
      & N^{m-2}(m+N-1)m \hspace{20pt} \text{if} \quad \ k = k' \\
      & N^{m-2}(m-1)m   \hspace{40pt} \text{if} \quad \ k \neq k' \, .
    \end{cases}
\label{eq:multinomial}
\end{align}
Note that, except for a different offset, Eq.~\eref{gn_2_foc_15} as a function of $\vec{r}_2$ displays the same peaked emission pattern as the result for $N$ SPE derived in \cite{Oppel(2014),Wiegner(2015)}; this has been commented for the case of two TLS already in Sec.~\ref{sec:HBT_2TLS}  (compare Eqs.~\eref{eq:g2} and \eref{eq:g2atoms}). 
The angular width of the central maximum (see Eq.~(\ref{gn_2_foc_15})) is given by
\begin{equation}
\label{peakwidth}
\delta\theta_2 \approx \frac{2\pi}{N\,k\,d}\ , 
\end{equation}
which is identical to the case $N$ SPE \cite{Oppel(2014),Wiegner(2015)}, i.e., displaying an increasingly peaked angular distribution for growing numbers of emitters $N$.

According to Eq.~(\ref{gn_2_foc_15}), the visibility of the $m$th-order correlation function for $N$ TLS is given by
\begin{equation}
\label{visibilityTLS}
{\cal V}_{TLS} = \frac{m-1}{m+1} \; ,
\end{equation}
i.e., independent of $N$. Moreover, it converges for high correlation orders $m\gg 1$ to $\mathcal{V}_{TLS} \approx 100 \%$.

Note further that when integrating $G_{N\,TLS}^{(m)}$ with respect to the phase $\varphi_{12}$ we obtain
\begin{equation}
\begin{aligned}
	\frac{1}{2\pi} \int &G_{N\,TLS}^{(m)}(\vec{r}_{1}, \hdots , \vec{r}_{1}, \varphi_{12})\, \text{d}\varphi_{12} \\
	&= \left< G_{N\,TLS}^{(m)} \right>_{\varphi_{12}} = \bar{n}^{m}N^{m}(m-1)!\frac{N+m-1}{N}\ .
\label{eq:integral}
\end{aligned}
\end{equation}
Hence for $m\gg N$ the normalized $m$th-order correlation function $G_{N\,TLS}^{(m)}/\left< G_{N\,TLS}^{(m)} \right>_{\varphi_{12}}$ has a central maximum which scales as $\sim N$, i.e., identical to the result for $N$ SPE in the case that $m=N$ \cite{Oppel(2014),Wiegner(2015)}. A contrast of $100 \%$, an angular width scaling as $\sim 1/N$ and a maximum value $\sim N$ are typical features of superradiance for symmetric Dicke states with one excitation \cite{Dicke(1954),Eberly(1971),Manassah(1973),Haroche(1982),AGARWAL(1974),Scully(2006),Scully(2007),Scully(2008),Scully(2009),Scully(2009)Super,Oppel(2014),Wiegner(2015),Scully(2015),Rohlsberger(2010),Rohlsberger(2013)}.

\begin{figure}[t!]
  \centering 
	\includegraphics[width=0.45\textwidth]{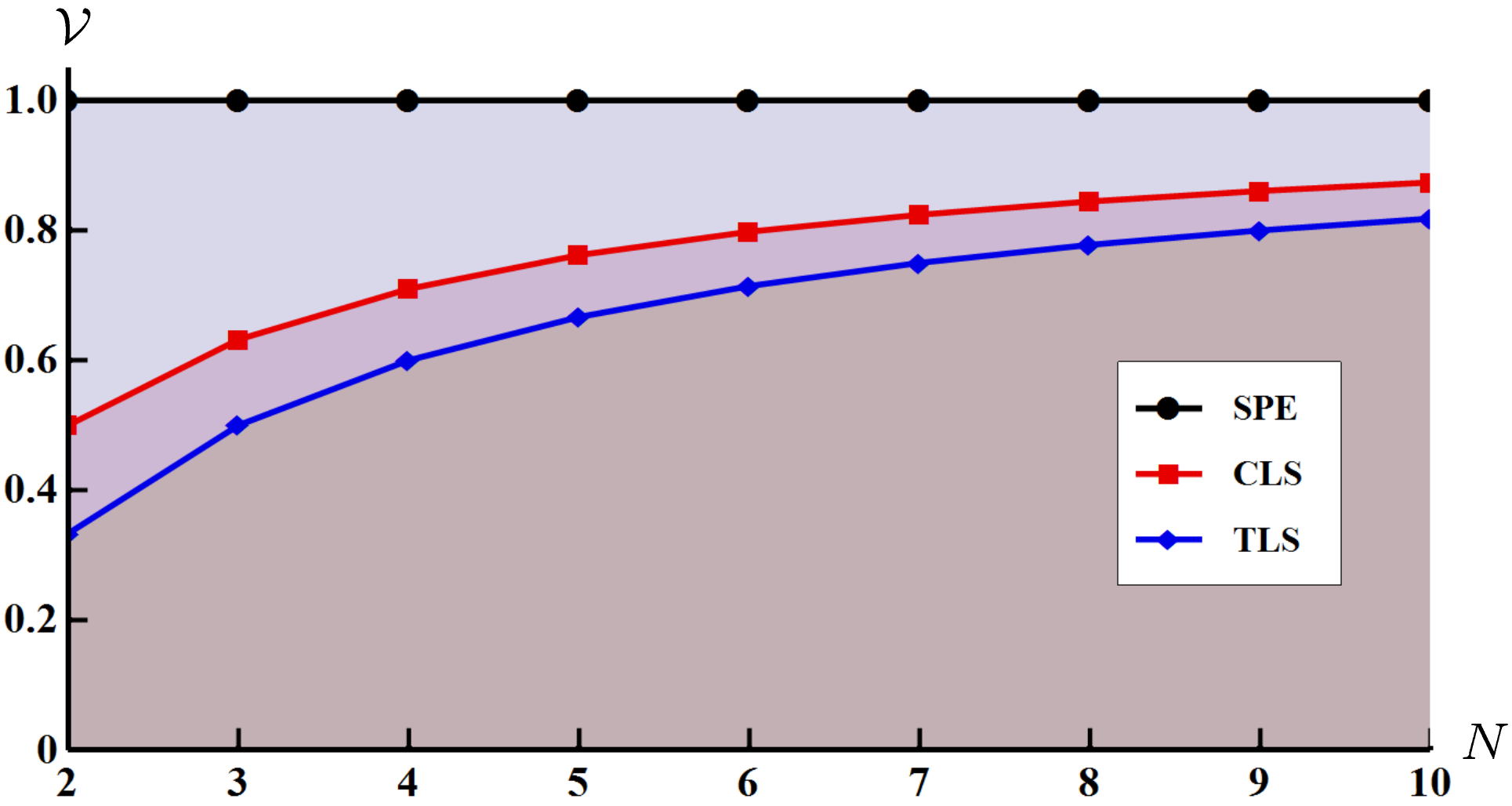}
  \caption{(Color online) Visibilities ${\cal V}$ of the $m$th-order correlation function $G^{(m)}_{N} (\vec{r}_1, ...,\vec{r}_1,\vec{r}_2)$ for $m = N = 2, \ldots, 10$ SPE (dots), CLS (squares) and TLS (rhombs), arranged along a chain at positions ${\vec R}_{l}$, $l = 1, \ldots, N$, with equal spacing $d >> \lambda$.}
  \label{fig:visibilitiy_gN_0}
\end{figure}

In the case of $N$ uncorrelated coherent light sources (CLS) a similar distribution as in Eq.~\eref{gn_2_foc_15} is obtained. Here we find (cf. Eq.~(\ref{eq:GNpartitions}))
\begin{equation}
\begin{aligned}
  \label{eq:GmNCLS}
  & \G{m}_{N\,CLS}(\vec{r}_1, ...,\vec{r}_1,\vec{r}_2)  \\
  & \hspace{2mm} = \frac{\bar{n}^{m}N}{m^{2}} \Big[ \sum_{m_{l}}\binom{m}{m_{1},\hdots,m_{N}}^{2} m_{k}m_{k} \\
	& \hspace{20mm} - \sum_{m_{l}}\binom{m}{m_{1},\hdots,m_{N}}^{2} m_{k}m_{k'} \Big] \\
	& \hspace{2mm} + \frac{\bar{n}^{m}}{m^{2}} \Big[ \sum_{m_{l}}\binom{m}{m_{1},\hdots,m_{N}}^{2} m_{k}m_{k'} \Big] \frac{\sin^2{\left(N\frac{\varphi_{11}-\varphi_{12}}{2}\right)}}{\sin^2{\left(\frac{\varphi_{11}-\varphi_{12}}{2}\right)}} \! ,
\end{aligned}
  \end{equation}
with $k\neq k'$.
Note that Eq.~\eref{eq:GmNCLS} displays the same angular width $\sim \frac{1}{N}$ as for $N$ TLS or $N$ SPE (cf. Eq.~(\ref{peakwidth})). Moreover, normalizing by integration for $m\gg N$ as in the case of TLS (cf. Eq.~(\ref{eq:integral})) gives the same superradiant scaling for the central maximum $\sim N$ as derived for $N$ TLS.
As concerns the visibility ${\cal V}_{CLS}$ we used Eq.~(\ref{eq:GmNCLS}) to calculate ${\cal V}_{CLS}$ for $m=N$ CLS numerically, since we were not able to simplify Eq.~(\ref{eq:GmNCLS}) further.
The  various visibilities obtained for $m= N = 2, \ldots, 10$ SPE, TLS and CLS are displayed in Fig.~\ref{fig:visibilitiy_gN_0}. It  shows that for $m=N$ ${\cal V}_{CLS}$ is always higher then ${\cal V}_{TLS}$, but lower than ${\cal V}_{SPE}$, independent of the correlation order $m$.

\section{Dicke like States for classical sources}
\label{sec:Dicke}

As we know that $\G{m}_{N\,SPE}(\vec{r}_1, ...,\vec{r}_1,\vec{r}_2)$ displays as a function of $\vec{r}_2$  the superradiant emission pattern radiated by $N$ SPE in the symmetric Dicke state $\ket{\frac{N}{2},\frac{N}{2}-(m-1)}$ (see \cite{Oppel(2014),Wiegner(2015)}), we conclude that $\G{m}_{N\,TLS}(\vec{r}_1, ...,\vec{r}_1,\vec{r}_2)$ 
as a function of $\vec{r}_2$ exhibits the superradiant emission characteristics of $N$ TLS being in an analogous Dicke state after $m-1$ photons have been recorded at $\vec{r}_1$. In the case of $N = m = 2$ this classical Dicke state has been given in Eq.~\eref{eq:rho_2TLS_proj}. For arbitrary numbers $N$ of TLS and arbitrary correlation order $m$ the corresponding \textit{classical Dicke state} reads 
\begin{equation}
\label{eq:thermal_state_matrice}
\tilde{\rho}_{N \, TLS}^{(m-1)}\sim (\sum_{l=1}^{N} e^{i\,\varphi_{l1}} \; \hat{a}_l)^{m-1} \rho_{N \, TLS} \, (\sum_{l=1}^{N} e^{-i\,\varphi_{l1}} \;\hat{a}^{\dagger}_l)^{m-1} \, .
\end{equation}
As already discussed for the case of $N=2$ TLS, we can see that Eq.~\eref{eq:thermal_state_matrice} is not of a diagonal form. The non-diagonal terms are due to the correlations between the $N$ classical sources induced by the measurement of $m-1$ photons at $\vec{r}_{1}$.
They are at the basis of the interference terms appearing in the subsequent intensity measurement at $\vec{r}_{2}$. 
Indeed, similar to Eqs.~\eref{eq:G2_rho_G1} - \eref{eq:rho_tilde} in Sec.~\ref{sec:HBT_2TLS}, we can rewrite Eq.~\eref{gn_2_foc_15} as
\begin{equation}
\begin{aligned}
	\G{m}_{N\,TLS}(\vec{r}_1, ...,\vec{r}_1,\vec{r}_2) & = G_{\tilde{\rho}_{N \, TLS}^{(m-1)}}^{(1)}(\vec{r}_{2}) G_{\rho_{N\, TLS}}^{(m-1)}(\vec{r}_{1},\hdots,\vec{r}_{1}) \ , 
\label{eq:Gm_rho_G1}
\end{aligned}
\end{equation}
where
\begin{equation}
G_{\tilde{\rho}_{N \, TLS}^{(m-1)}}^{(1)}(\vec{r}_{2}) = \text{Tr}\left[\tilde{\rho}_{N \, TLS}^{(m-1)}  E^{(-)}(\vec{r}_{2})E^{(+)}(\vec{r}_{2})\right] \ ,
\label{eq:G_rho_m_tilde}
\end{equation}
with $\tilde{\rho}_{N \, TLS}^{(m-1)}$ given in Eq.~\eref{eq:thermal_state_matrice}. 
Eqs.~\eref{eq:Gm_rho_G1} and \eref{eq:G_rho_m_tilde} show that for arbitrary classical fields $\rho_{N \, TLS}$ with arbitrary number of TLS we find again the isomorphism between $G_{N \, TLS}^{(m)}(\vec{r}_{1}, \ldots, \vec{r}_{1}, \vec{r}_{2})$ (Eqs.~\eref{gn_2_foc_15} and \eref{eq:Gm_rho_G1}), and the intensity measurement $G_{\tilde{\rho}_{N \, TLS}^{(m-1)}}^{(1)}(\vec{r}_{2})$ of the projected field $\tilde{\rho}_{N \, TLS}^{(m-1)}$ (Eq.~\eref{eq:G_rho_m_tilde}).
Note that the proportionality factor in Eq.~\eref{eq:thermal_state_matrice} can be derived from
\begin{equation}
\text{Tr}[\tilde{\rho}_{N \, TLS}^{(m-1)}] = 1 \, . 
\label{eq:rho_m_tilde-Trace}
\end{equation}

In Eqs.~\eref{gn_2_foc_15} and \eref{eq:Gm_rho_G1}, the case $m = 2$ corresponds to the celebrated Hanbury Brown and Twiss experiment \cite{HBT(1956)Correlation,Brown(1967),HANBURY(1974)}. Indeed, according to the discussion above, the experiment conducted in 1956 by Hanbury Brown and Twiss can be reinterpreted as the first measurement of superradiance with classical sources:
Taking Eq.~(\ref{gn_2_foc_15}) in the limit $N \rightarrow \infty$ such that $N d = D$ equals the diameter of the star, we find for $m=2$ and $\theta_1(\vec{r}_1) = 0$
\begin{equation}
\begin{aligned}
  \label{gn_2_foc_15_m2}
  & \g{2}_{N\,TLS}(\vec{r}_1,\vec{r}_2) = \left( 1 + \text{sinc}^2 \left(\frac{D k \sin \theta_{2}}{2}\right)\right) \, ,
\end{aligned}
\end{equation}
corresponding to the well-known results reported by Hanbury Brown and Twiss in \cite{HBT(1956)Correlation}.

\section{Compact way for calculating higher-order intensity correlations}
\label{sec:functional}

In this section we present a different way to calculate intensity correlations of arbitrary order $G^{(m)}_{N}(\vec{r}_{1},\vec{r}_{2},\hdots,\vec{r}_{m})$ compared to Secs.~\ref{sec:HBT_2TLS} - \ref{sec:Dicke}, based on the characteristic functional $C[f(\cdot)]$. In this approach the different orders of the correlation functions can be obtained by mere differentiation.

The characteristic functional $C[f(\cdot)]$ can be defined via \cite{KLAUDER(1968)}
\begin{equation}
\begin{aligned}
	C[F(\cdot)] = &\left<\exp\left\{i\int F^{*}(\vec{r}) E^{(-)}(\vec{r})\text{d}^{3}r \right\}\right. \\
	&\hspace{15mm} \times \left.\exp\left\{i\int F(\vec{r}) E^{(+)}(\vec{r})\text{d}^{3}r \right\} \right> \, ,
\label{eq:FunctionalDef}
\end{aligned}
\end{equation}
where $F(\vec{r})$ is an arbitrary function and $E^{(+)}(\vec{r})$ ($E^{(-)}(\vec{r})$) is the positive (negative) frequency part of the electric field at position $\vec{r}$, given by the sum over all field contributions of the distinct sources
\begin{equation}
	E^{(\pm)}(\vec{r})=\sum_{j=1}^{N}E^{(\pm)}_{j}(\vec{r}) \ .
\label{eq:definitionfieldthermal}
\end{equation}
From Eqs.~\eref{eq:FunctionalDef} and \eref{eq:definitionfieldthermal} we immediately obtain the general form of $G^{(m)}_{N}(\vec{r}_{1},\vec{r}_{2},\hdots,\vec{r}_{m})$, valid for arbitrary light fields
\begin{equation}
\begin{aligned}
	&G^{(m)}_{N}(\vec{r}_{1},\vec{r}_{2},\hdots,\vec{r}_{m}) \\
	&\!\!= \!\! \left. (-1)^{m} \frac{\delta^{2m}C[F(\cdot)]}{\delta F^{*}(\vec{r}_{1})\hdots \delta F^{*}(\vec{r}_{m})\delta F(\vec{r}_{1})\hdots \delta F(\vec{r}_{m})} \right|_{\substack{F(\vec{r}_{i})=0 \\
	F^{*}(\vec{r}_{j})=0}} \! ,
\label{eq:GmNFunctional}
\end{aligned}	
\end{equation}
with $i,j=1,\hdots,m$. Note that the definiton of $C[F(\cdot)]$ in Eq.~\eref{eq:FunctionalDef} is valid for both quantum and classical light fields.

In the case of $N$ TLS the characteristic functional, Eq.~(\ref{eq:FunctionalDef}), can be written in the form
\begin{equation}
\begin{aligned}
	&C[F(\cdot)]\\
	&=\prod_{j=1}^{N}\left<\exp\left\{i\int F^{*}(\vec{r}) E_{j}^{(-)}(\vec{r}) \text{d}^{3}r \right\} \right. \\
	& \hspace{25mm} \times \left. \exp\left\{i\int F(\vec{r}) E^{(+)}_{j}(\vec{r}) \text{d}^{3}r \right\} \right> \\
		&= \prod_{j=1}^{N}\exp\left\{-\iint \text{d}^{3}r \; \text{d}^{3}r' \; F^{*}(\vec{r}) \; F(\vec{r}') \; \Gamma_{j}(\vec{r}-\vec{r}') \right\} \\
		&= \exp\left\{-\iint \text{d}^{3}r \; \text{d}^{3}r' \; F^{*}(\vec{r}) \; F(\vec{r}') \; \sum_{j=1}^{N} \Gamma_{j}(\vec{r}-\vec{r}') \right\} \ ,
\label{eq:CTLS}
\end{aligned}
\end{equation}
with $\Gamma_{j}(\vec{r}-\vec{r}') = \left<E_{j}^{(-)}(\vec{r})E^{(+)}_{j}(\vec{r}')\right>$. Now using Eq.~(\ref{eq:CTLS}) in Eq.~(\ref{eq:GmNFunctional}) one immediately obtains the Gaussian moment theorem involving only first moments of $E^{(\pm)}(\vec{r})$, which, for $\vec{r}_{1}=\hdots=\vec{r}_{m-1}$, can be easily calculated. This  leads to the explicit form of $G^{(m)}_{N}(\vec{r}_{1},\ldots,\vec{r}_{1},\vec{r}_{2})$ given by Eq.~(\ref{gn_2_foc_15}). Note that by use of Eq.~(\ref{eq:GmNFunctional}) one can also show that the Gaussian moment theorem is identical to Wick's theorem.

To calculate $G^{(m)}_{N \, CLS}(\vec{r}_{1},\ldots,\vec{r}_{1},\vec{r}_{2})$ for $N$ CLS one can employ a discrete version of Eq.~(\ref{eq:GmNFunctional})
\begin{equation}
\label{eq:C_F_dot}
\begin{aligned}
	C[F(\cdot)] =\prod_{l=1}^{N} \left< e^{a_{l}^{\dagger}\beta_{l}^{*}} e^{a_{l}\beta_{l}}  \right> \, .
\end{aligned}
\end{equation}
In the superradiant case we have
\begin{equation}
	\beta_{j}= c_{1l} f_{1} + c_{2l} f_{2} \, ,
\end{equation}
where $f_{j}$ and $f^{*}_{j}$ are functionals and $c_{lj}=\exp\left\{i k\, \vec{n}_{j}\cdot \vec{R}_{l}\right\}$ describes the phase accumulated by a photon traveling from source $l$ to detector $j$ (cf. Eq.~\ref{eq:Eplus}). For coherent fields it can be shown that
\begin{equation}
\label{eq:C_F}
\begin{aligned}
	C[F(\cdot)] =\prod_{l=1}^{N} J_{0}(2|\alpha| |\beta_{l}|) \ ,
\end{aligned}
\end{equation}
with $J_{0}$ being the Bessel function of first kind and $\alpha=\alpha_{l}$ the complex amplitude of the $l$th CLS.

Plugging Eq.~\eref{eq:C_F} into Eq.~\eref{eq:C_F_dot} the functional approach leads to
\begin{equation}
\begin{aligned}
&G^{(m)}_{N\ CLS}(\vec{r}_{1},\ldots,\vec{r}_{1},\vec{r}_{2}) =  \frac{(-1)^{m}}{N-1}  \times \\
& \hspace{2mm}  \left[ |\alpha|^{2} \left(\left.\Delta^{m-1} J_{0}^{N}(2|\alpha| |f_{1}|)\right|_{\substack{f_{1}=0 \\	f_{1}^{*}=0}}\right)  \right. \\ 
& \hspace{25mm} \times \Big[ -N^{2} + \frac{\sin^2{\left(N\frac{\varphi_{11}-\varphi_{12}}{2}\right)}}{\sin^2{\left(\frac{\varphi_{11}-\varphi_{12}}{2}\right)}} \Big]  \\
& \hspace{7mm} +  \left(\left.\Delta^{m} J_{0}^{N}(2|\alpha| |f_{1}|)\right|_{\substack{f_{1}=0 \\	f_{1}^{*}=0}}\right) \\
& \hspace{25mm} \left. \times \Big[ - 1 + \frac{1}{N}\frac{\sin^2{\left(N\frac{\varphi_{11}-\varphi_{12}}{2}\right)}}{\sin^2{\left(\frac{\varphi_{11}-\varphi_{12}}{2}\right)}} \Big] \right]   ,
\label{eq:GmNCLS2}
\end{aligned}
\end{equation}
with $\Delta=\delta^{2}/(\delta f_{1} \, \delta f^{*}_{1})$ and the mean photon number of a single source $|\alpha|^{2}=\bar{n}$. The identity of Eq.~(\ref{eq:GmNCLS}) and Eq.~(\ref{eq:GmNCLS2}) can be proven 
by use of \cite{Moll(2014)}
\begin{equation}
\begin{aligned}
	\left.\Delta^{m} J_{0}^{N}(2|\alpha| |f_{1}|)\right|_{\substack{f_{1}=0 \\	f_{1}^{*}=0}} & = (-1)^{m} |\alpha|^{2m} B_{m}^{(0)}(N) \\
	& = (-1)^{m} |\alpha|^{2m} W_{N}^{(0)}(2m) \ ,
\end{aligned}
	\label{eq:B_m_N}
\end{equation}
where the following recurrence relation for $B_{m}^{(0)}(N)$ holds
\begin{equation}
	B_{m}^{(0)}(N) = \sum_{k=1}^{m} [\frac{k(N+1)}{m}-1]\binom{m}{k}^{2} B_{m-k}^{(0)}(N) \ ,
\end{equation}
with $B_{0}^{(0)}(N)=1$. Note that the second possible solution $W_{N}^{(0)}(2m)$ in Eq.~\eref{eq:B_m_N} is known from the uniform theory of random walks \cite{Borwein(2011)}
\begin{equation}
\begin{aligned}
	W_{N}^{(0)}(2m) &= \int_{[0,1]^{N}} \left| \sum_{k=1}^{N}e^{i 2\pi x_{k}} \right|^{2m} d\vec{x} \\
	& = \sum_{m_{l}}\binom{m}{m_{1},\hdots,m_{N}}^{2} \ ,
\end{aligned}
\end{equation}
\\
\noindent describing the $2m$th moment of the distance to the origin after $N$ steps. This moment exactly yields the value of the central peak of the $m$th-order correlation function, Eq.~(\ref{eq:GmNCLS}), when all $m$ detectors are placed at the same position.

\section{Conclusion}
\label{sec:conclusion}

In this paper we discussed the various aspects of spatial superradiance by distilling its prerequisites and fundamental properties in a rigorous mathematical way. In particular, we showed that a peaked superradiant angular distribution of the emitted radiation can not only be displayed by quantum emitters prepared in highly entangled Dicke states but equally by classical sources prepared in corresponding \textit{classical Dicke states} via projective measurements of photons in the far field of the sources. Here, upon detection of the last photon the same permutative superposition of quantum paths appear in the $m$th-order intensity correlation function $G^{(m)}_{N} (\vec{r}_1, \ldots, \vec{r}_1, \vec{r}_2)$ than those producing superradiance in the case of initially uncorrelated single photon emitters (SPE). The difference is that for SPE only the quantum paths of the single partition $\{1, 1, \ldots, 1\}$ add to $G^{(m)}_{N} (\vec{r}_1, \ldots, \vec{r}_1, \vec{r}_2)$ whereas in the case of classical sources the quantum paths \textit{all} possible partitions of the number $m$ contribute. The latter causes a slightly reduced visibility of the superradiant angular distribution compared to the one produced by SPE. Yet, for increasing numbers of recorded photons $m\gg N$ the visibility converges to $\mathcal{V}=100 \, \%$ also in the case of classical sources.

The analysis shows in particular that, equally to the case of quantum sources, the state of a classical system can be manipulated by recording photons in the far field of the sources such that the particular photon source remains unknown. This is a further example of the production of correlations among classical sources due to a measurement process \cite{Oppel(2012)}.

The arrangement required to display the superradiant spatial emission pattern of classical sources corresponds to a generalized Hanbury Brown and Twiss setup correlating photons at different positions emitted from the initially uncorrelated sources. In this way we show that, similar to the case of SPE discussed in a foregoing paper \cite{Wiegner(2015)}, it is possible to employ statistically independent and initially uncorrelated classical sources to produce a superradiant peaked angular distribution for the last emitted photon. In particular, we demonstrate that for thermal light sources (TLS) the celebrated Hanbury Brown and Twiss effect, originally established in astronomy to determine the dimensions or distances of stars \cite{HBT(1956)Correlation,Brown(1967),HANBURY(1974)}, and Dicke superradiance, commonly observed with atoms in symmetric Dicke states \cite{Dicke(1954),Eberly(1971),Manassah(1973),Haroche(1982),AGARWAL(1974)}, are two sides of the same coin.

\section{Acknowledgement}

The authors gratefully acknowledge funding by the Erlangen Graduate School in Advanced Optical Technologies (SAOT) by the German Research Foundation (DFG) in the framework of the German excellence initiative. D.B. gratefully acknowledges financial support by the Cusanuswerk, Bisch\"ofliche Studienf\"orderung. R.W. and S.O. gratefully acknowledge financial support by the Elite Network of Bavaria and the hospitality at the Oklahoma State University. This work was supported by the DFG research grant ZA 293/4-1. G.S.A. is especially grateful to the SAOT for providing financial grant for making this collaboration possible.

\bibliography{literatur_superradiance_II}
\bibliographystyle{apsrev4-1}

\end{document}